\documentstyle[twoside,fleqn,espcrc2,epsf]{article}

\newcommand{\be}{\begin{equation}}
\newcommand{\ee}{\end{equation}}

\newcommand{\AmS}{{\protect\the\textfont2
  A\kern-.1667em\lower.5ex\hbox{M}\kern-.125emS}}

\title{Semileptonic $B$ Decays with $N_f = 2+1$ Dynamical Quarks}

\author{J.Shigemitsu\address{Physics Department, The Ohio State
        University, Columbus, OH 43210, USA.},
        C.T.H.Davies$^{\rm b}$, A.Dougall\address{Department of Physics \&
               Astronomy, University of Glasgow, Glasgow, G12 8QQ, UK.},
        K.Foley\address{Laboratory of Elementary Particle Physics,
        Cornell University, Ithaca, NY 14853, USA.}, 
        E.Gamiz$^{\rm b}$, A.Gray$^{\rm a}$,  
        E.Gulez$^{\rm a}$, 
        G.P.Lepage$^{\rm c}$,
        M.Wingate\address{Institute for Nuclear Theory, University of
        Washington, Seattle, WA 98115, USA.}
              }

\begin{document}

\begin{abstract}
 Semileptonic, $B \rightarrow \pi  \, l \overline{\nu}$,
 decays are studied on the MILC dynamical configurations
 using NRQCD heavy and Asqtad light quarks.
 We work with light valence quark masses
ranging between $m_s$ and $m_s/8$. Preliminary simple linear
chiral extrapolations have been carried out for form factors $f_\parallel$
and $f_\perp$ at fixed $E_\pi$. The chirally extrapolated results
for the form factors $f_+(q^2)$ and
$f_0(q^2)$ are then fit to the Becirevic-Kaidalov (BK) ansatz. 
Preliminary estimates of the CKM matrix element $|V_{ub}|$ are 
presented based on recently published branching fractions 
for  $B^0 \rightarrow \pi^-  \, l^+ \nu$ exclusive decays 
by the CLEO collaboration.
\vspace{1pc}
\end{abstract}

\maketitle

\section{Introduction}
First principles calculations of $B$ meson semileptonic decay form 
factors are crucial for determining the CKM matrix elements $|V_{cb}|$ 
and $|V_{ub}|$. Recent progress on the lattice towards this goal comes 
from two major developments: the ability to go beyond the quenched 
approximation with close to realistic dynamical quark content \cite{milc,prl} 
and the use of improved staggered light quarks in 
heavy-light simulations \cite{matt}.
  We report here on unquenched studies of 
$B \rightarrow \pi, \, l \overline{\nu}$  decays on the lattice 
using one of the coarse MILC $N_f = 2+ 1$ dynamical sets
 \cite{milc}, NRQCD $b$ quarks 
 and  improved staggered 
(Asqtad) light quarks.
The light dynamical quark mass is fixed at $m_{dyn} = m_s/4$  and
we vary the light valence quark mass between $m_s$ and $m_s/8$.

\section{ Form Factors}

Semileptonic form factors parameterise the hadronic matrix elements
of electroweak currents 
between a $B$ meson and a $\pi$ or a $\rho$. In particular, one has
\begin{eqnarray}
 \langle \pi| V^\mu| B \rangle  &=&
 f_+(q^2) 
\, \left[ p_B^\mu + p_\pi^\mu - \frac{M_B^2-m_\pi^2}{q^2}
\, q^\mu \right] \nonumber \\
 && \; + \;
f_0(q^2)
 \, \frac{M_B^2 - m_\pi^2}{q^2} \, q^\mu  \\
 &=& \sqrt{2 M_B} \,[v^\mu f_\parallel\,
 + \, p^\mu_\perp  f_\perp ]  
\end{eqnarray}
with \\
 $v^\mu = \frac{p_B^\mu}{M_B}$,  $\; p_\perp^\mu
= p_\pi^\mu - (p_\pi \cdot v) \, v^\mu$, $\; q^\mu = p_B^\mu - p_\pi^\mu$.

\vspace{.1in}
\noindent
A lattice calculation of the relevant matrix element starts with 
the three-point correlator 
\begin{eqnarray}
& & C^{(3)}(\vec{p}_\pi, \vec{p}_B, t, T_B)   =   \nonumber \\  
 & & \sum_{\vec{z}} \sum_{\vec{y}} \langle \; \Phi_\pi(0) \;
V^\mu_{lat}(\vec{z},t) \; \Phi^\dagger_B(\vec{y},T_B) \; \rangle \nonumber \\
& & \qquad \qquad \times 
\, e^{i\vec{p}_B\cdot \vec{y}} \, e^{i(\vec{p}_\pi - \vec{p}_B)\cdot \vec{z}}
\end{eqnarray}
where $\Phi_\pi$ and $\Phi_B$ are interpolating operators for the
$\pi$ and $B$ mesons respectively.
The three-point correlator is fit to the form 
\begin{eqnarray}
&& C^{(3)}(\vec{p}_\pi, \vec{p}_B, t, T_B)   \rightarrow 
  \nonumber \\  
 & &  \sum_{k=0}^{N_\pi-1} \sum_{j=0}^{N_B-1}
 (-1)^{k*(t-1)} \, (-1)^{j*(T_B-t)} 
 \qquad \qquad \qquad \nonumber \\
&& \qquad \quad \times  A_{j,k}
\,  e^{-E_\pi^{(k)} (t-1)} \, e^{ -E_B^{(j)} (T_B-t)}
\end{eqnarray}
and the goal is to extract the ground state contribution $A_{00}$.
Working in the $B$ meson rest frame, the ground state amplitudes 
$A_{00}(V^\mu)$ are related to the form factors $f_\parallel$ 
and $f_\perp$ in a simple way.
\begin{eqnarray}
f_{\|} & = & \frac{ A_{00}(V^0)}
{\sqrt{\zeta^{(0)}_\pi \zeta^{(0)}_B}}
\, \sqrt{2 E_\pi} \, Z_{V_0}   \\
f_{\perp} & = & \frac{ A_{00}(V^k)}
{\sqrt{\zeta^{(0)}_\pi \zeta^{(0)}_B}
\;p_\pi^k }
\, \sqrt{2 E_\pi} \, Z_{V_k}  
\end{eqnarray}
where $\zeta^{(0)}_{\pi,B}$ are the groundstate amplitudes of 
the $\pi$ and $B$ two-point correlators respectively and $Z_{V_\mu}$ 
are the matching factors between the lattice heavy-light current 
$V^\mu_{lat}$ and the current in the continuum theory. We use one-loop
estimates for these matching factors \cite{emel}.

\begin{figure}
\epsfxsize=7.5cm
\centerline{\epsfbox{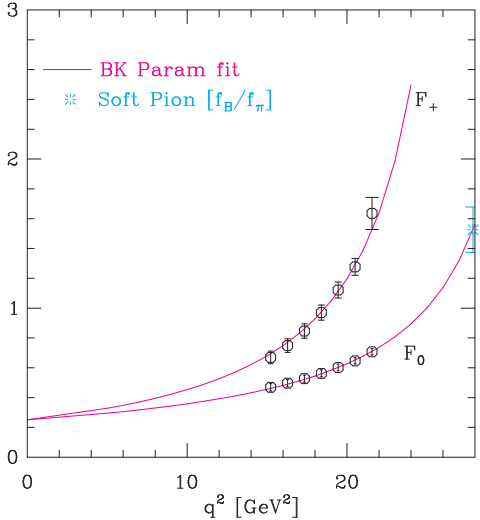}}
\caption{The form factors $f_+(q^2)$ and $f_0(q^2)$, extrapolated 
to the physical pion.  Only statistical errors are shown. 
The full curves correspond to a Becirevic-Kaidalov (BK) 
parametrization fit to the data (see text).
 }
\end{figure}

\section{ The Form Factors $f_+(q^2)$ and $f_0(q^2)$ at the Physical Pion }
Although our simulations have been carried out with light quark masses 
as low as $m_s/8$ one still needs to extrapolate the form factors, 
determined above, to the physical pion.  To date, we have only carried out 
simple linear chiral extrapolations.
 We first interpolate $f_\parallel$ and $f_\perp$
 to common values of $E_\pi$, the pion energy in the $B$ rest frame. These are 
then extrapolated linearly to the physical pion for several fixed values of 
$E_\pi$. 

From the chirally extrapolated $f_\parallel$ and $f_\perp$ one 
obtains the form factors $f_+(q^2)$ and $f_0(q^2)$ at the physical 
pion.  This is shown as the circles in Fig.1. Our results are currently 
limited to the $q^2 \geq 15\,GeV^2$ region. 
 Recently, a very promising 
approach to low $q^2$ form factors has been developed, namely 
``Moving NRQCD'', 
which will allow us to overcome this limitation \cite{kfoley,mnrqcd}. 
 In the mean time, however,
we will rely on a model ansatz to extend our form factor results into 
the low $q^2$ regime. Specifically, we employ an ansatz introduced by 
Becirevic \& Kaidalov (BK) \cite{bk},
\begin{eqnarray}
f_+(q^2) &=& \frac{C_B \, (1- \alpha_B)}
{(1-\tilde{q}^2)(1-\alpha_B
\tilde{q}^2)} \\
f_0(q^2) &=& \frac{C_B \, (1- \alpha_B)}{(1-\tilde{q}^2/\beta_B)}
\end{eqnarray}
 $(\tilde{q}^2 \equiv q^2/M_{B^*}^2)$.  This ansatz satisfies the 
kinematic constraint $f_+(0) = f_0(0)$, HQET scaling laws and the 
requirement of a pole in $f_+(q^2)$ at $q^2 = M_{B^*}^2$. 
We find an excellent fit to this BK ansatz using the physical $M_{B^*}$ 
mass and this is shown as the full curves in Fig.1 
(a satisfactory BK parametrization was not possible before the 
chiral extrapolation).  The fit parameters are 
$$C_B = 0.42(3), \;\; \alpha_B = 0.41(7), \;\;
\beta_B = 1.18(5),$$
which translates into 
$$f_0(0) = f_+(0) = 0.251(15) $$
and an effective pole in $f_0(q^2)$ at $ q^2 = 
 (M^{pole}_{f0})^2 = 33.35(1.36$)GeV$^2 $.  Both $f_{0,+}(0)$ and 
$M^{pole}_{f0}$ are in good agreement with a recent semileptonic $B$ decay 
analysis based on Sum Rules \cite{ball}.
The data points in Fig.1 and the $f_{0,+}(0)$ 
quoted above include only statistical
and fitting errors.  Further systematic errors are discussed in 
the next section. A comparison of our new dynamical form factor results 
with old quenched data is given, for instance, in reference \cite{alan}.

\section{ Estimating $|V_{ub}|$ }
The CLEO collaboration has published branching fractions
for exclusive semileptonic $B$ decays, including binning into 
several $q^2$ ranges \cite{cleo}.  We combine 
these experimental inputs with lattice 
results for $f_+(q^2)$ to extract values for the CKM matrix element 
$|V_{ub}|$.  The differential decay rate for $B^0 \rightarrow \pi^-, \, l^+
 \nu$,
\be
\frac{1}{|V_{ub}|^2} \, \frac{d\Gamma}{dq^2} = \frac{G_F^2 }
{24 \pi^3 } \, p_\pi^3 \, |f_+(q^2)|^2 
\ee
can be integrated to give 
$\frac{\Gamma}{|V_{ub}|^2}$ and 
the partial width $\Gamma$ can be determined 
from CLEO's branching fraction and the 
Particle Data Group's $B^0$ lifetime of $1.542 \pm 0.016 \, ps$.
Our \underline{ preliminary} estimate for $|V_{ub}|$ is then 
$$ |V_{ub}| = \left\{ \begin{array}{l}
 3.86(32)(58) \times 10^{-3}
 \qquad 0 \leq q^2 \leq q^2_{max}     \\
                                       \\
 3.52(73)(44) \times 10^{-3} \qquad  16 \, GeV^2 \leq q^2 \\
                          \end{array} \right.
                                              $$
where the two values correspond to either using the entire allowed
$q^2$ range or restricting both experiment and theory to the 
$q^2 \geq 16$ GeV$^2$ region.  The first error is experimental and 
the second is our current best estimate of lattice statistical 
and systematic errors added in quadrature. In addition to $4 \sim 6$\% 
statistical errors, we estimate  $\sim 9$\% higher order 
perturbative matching, $\sim 5$\% chiral extrapolation, $\sim 5$\% 
relativistic and discretization errors. This adds up to $\sim12.5$\% 
lattice errors for $|V_{ub}|$ obtained from the $q^2 \geq 16$ GeV$^2$ region.
For $|V_{ub}|$ based on the full $q^2$ range we increase the lattice 
errors to $15$\% (an additional $8$\% added in quadrature) 
taking into account the need to rely on the BK 
parametrization to enter the low $q^2$ region.

\section{Summary }
Unquenched simulations of heavy meson semileptonic decays are now feasible and 
we report here on the first such calculations using NRQCD heavy and 
Asqtad light quarks (see talk by Okamoto for results using Fermilab 
heavy quarks \cite{okamoto}). The use of the improved staggered light
quark action has allowed for
significantly smaller statistical  and chiral extrapolation errors 
than in the past.
Combining lattice results for $f_+(q^2)$ with experimental branching 
fraction data has led to preliminary estimates of $|V_{ub}|$.\\
Many improvements are planned: inclusion of all dimension 4
($1/M$,  $\alpha_s/M$ and $a \alpha_s$) current corrections, more 
sophisticated chiral extrapolations \cite{scpt}, and simulations 
at other dynamical quark masses and lattice spacings. Use of ``Moving 
NRQCD'' \cite{kfoley} will also allow us to simulate directly at lower $q^2$.

\vspace{.1in}
\noindent
Acknowledgements : This work was supported by the DOE, 
 PPARC and NSF. Simulations were carried out 
 at NERSC.

\end{document}